\documentclass[doublespacing]{elsart}

\usepackage{epsfig}

\usepackage{amssymb}

\begin{document}

\begin{frontmatter}

\title{Dephasing and thermal smearing in an electromechanical which-path device}

\author[label1]{A. D. Armour\corauthref{cor1},}
\author[label2]{M. P. Blencowe}
\address[label1]{Blackett Laboratory, Imperial College, London SW7 2BW, U.K.}
\address[label2]{Department of Physics and Astronomy, Dartmouth College,
Hanover, New Hampshire 03755}
\corauth[cor1]{Corresponding author:  Fax: +44-(0)-20-7594-7604,
email: a.armour@ic.ac.uk
}

\begin{abstract}
In an electromechanical which-path device electrons travelling through an 
Aharonov--Bohm  ring with a quantum dot in one of the arms are dephased by an
interaction with the fundamental flexural mode of a radio-frequency cantilever,
leading to a reduction in the visibility of interference fringes. However,
at finite temperatures time-averaged measurement of the current 
leads to a fringe visibility which is reduced partly by dephasing of the
electrons and partly by a thermal smearing effect. The balance between
the thermal smearing and dephasing predicted  by a calculation depends
very strongly on the choice of cantilever basis states used. 
The interaction between the cantilever and its 
environment is expected to select the coherent state basis for the cantilever
and hence lead to a dephasing rate which is substantially lower than that which 
would arise if instead the  Fock states were selected.  
\end{abstract}

\end{frontmatter}
 
In a which-path experiment an external measuring device is used to probe 
for the presence of a particle in one of two interfering paths which the 
particle can take. Complementarity requires that any knowledge gained about
the path taken by a particle must  be  accompanied by a loss of interference
fringe visibility, thus which-path systems provide an ideal testing ground
for theoretical ideas about dephasing.
In the solid-state an Aharonov-Bohm (AB) ring may be used to produce interfering 
paths for electrons; which-path measurements can then be  performed by 
incorporating a quantum dot on one arm of the ring and introducing an 
external device which couples to the occupancy of the dot (i.e. by an 
electrostatic interaction). Buks {\emph{et al.}}\cite{buks} 
demonstrated a solid-state which-path device by using the current through a 
quantum point contact adjacent to the dot to probe the dot's occupancy.
 
Recently, we developed a theoretical model to describe an electromechanical
version of the which-path experiment\cite{ab}. We considered a system 
consisting of an AB ring with a quantum dot on one arm, similar to that used by
Buks {\emph{et al.}}, but with a cantilever positioned so that its tip lies just 
above the dot as opposed to a quantum point contact. If the cantilever is at 
a high enough voltage with respect to the dot then it can act as a which-path
detector: the presence (and absence) of the additional electron on the dot
in the Coulomb blockade regime couples to the flexural modes of the
cantilever. It turns out that at sufficiently low temperatures only the 
lowest flexural mode of the cantilever is relevant and so it can be 
treated as a single-mode quantum oscillator.  

In the theory, we assumed that the cantilever was initially
in a thermal mixture of states so that the overall fringe visibility is 
obtained by summing over the elastic transmission for a complete set of 
cantilever states\cite{ab}. The reduction in fringe visibility which we obtain
from the thermally weighted average is composed of two parts: the magnitudes of 
the elastic transmission for each cantilever state are reduced by the dephasing 
caused by the cantilever and the phase of the transmission differs for each 
cantilever state. When the average is taken the phase differences cause a further 
reduction in magnitude (an effect which we term thermal smearing). Thus in 
calculating the thermally averaged transmission, we can predict the 
time-independent fringe visibility, but we are not able to predict how much 
dephasing the cantilever causes. Here we investigate the distinction
between contributions to the reduction in interference fringe visibility 
from the cantilever-induced dephasing of the electron and thermal smearing.
 We calculate the elastic transmission for the cantilever initially in a pure 
 Fock state, as well as the case where the cantilever is in a pure coherent 
 state which we considered before. The dephasing induced by the cantilever is 
 strongly dependent on the state it is in, but a thermal average leads to a 
 unique result and so does not of itself give complete information about 
 the dephasing induced by the cantilever. 

The key quantity in the theory is the elastic transmission amplitude
of the arm containing the dot, $\langle t_{QD}(\epsilon)\rangle$, for a given
electron energy in the leads, $\epsilon$, since it determines the visibility of
the interference fringes.  For a cantilever in the pure state 
$|a\rangle$, the elastic transmission takes the form
\[
\langle a| t_{\mathrm QD}(\epsilon)| a\rangle=-\Gamma
\int_0^{\infty}\frac{d\tau}{\hbar}
{\mathrm{e}}^{[i(\epsilon-\epsilon_0)-\Gamma/2]\tau/\hbar}\langle
{\Phi}(\tau)\rangle, 
\]
where $\hbar/\Gamma$ is the characteristic dwell time of the electron on
 the dot, $\epsilon$ and $\epsilon_0$ are the energies of the electrons 
 incident on the dot and the relevant energy level in the dot respectively, 
 and the cantilever dependent term is
\[
\langle{\Phi}(t)\rangle=\langle a|\mathrm{e}^{i\omega c^{\dagger}c t}
\mathrm{e}^{-i\omega c^{\dagger}c t +i\lambda(c^{\dagger}+c)t/\hbar}| a\rangle
\]
where $c$ operates on the fundamental flexural mode of the cantilever with
frequency $\omega$ and $\lambda$ is the strength of the (linear) 
coupling between the cantilever and the electron on the dot.

The value of the elastic transmission is strongly dependent on the state of 
the cantilever. For a pure coherent state, $|\nu\rangle$, the cantilever 
dependent part of the transmission, $\langle{\Phi}(t)\rangle$, takes the 
form\cite{ab},
\[
\langle \nu |{\Phi}(t)| \nu \rangle=
{\mathrm{e}}^{i(\lambda/\hbar\omega_0)^2[\omega_0\tau-\sin(\omega_0 \tau)]}
{\mathrm{e}}^{-(\lambda/\hbar\omega_0)^2[1-\cos(\omega_0\tau)]}
{\mathrm{e}}^
{(\lambda/\hbar\omega_0)(\nu^*\mu-\nu\mu^*)},
\]
where $\mu={\mathrm{e}}^{i\omega_0 t}-1$.
In contrast, if the cantilever is in a Fock state $| n\rangle$ then 
$\langle{\Phi}(t)\rangle$ is given by\cite{Mahan}
\begin{eqnarray*}
\langle n|{\Phi}(t)| n\rangle&=&
{\mathrm{e}}^{i(\lambda/\hbar\omega_0)^2[\omega_0\tau-\sin(\omega_0 \tau)]}
{\mathrm{e}}^{-(\lambda/\hbar\omega_0)^2[1-\cos(\omega_0\tau)]}\\
&&\times L_n(2(\lambda/\hbar\omega_0)^2[1-\cos(\omega_0\tau)]),
\end{eqnarray*}
where $L_n(x)$ is a Laguerre polynomial of order $n$. 

For given values of the electron energies and lifetimes on the dot, the elastic
transmission is a function of the cantilever's state and the strength of the 
cantilever-dot coupling, $\lambda$.  When there is a finite coupling between 
the cantilever and the dot, the transmission amplitude is 
reduced in magnitude and receives  a phase shift (i.e. a finite argument). 
The reduction in magnitude is readily interpreted as dephasing of the 
electron, but it can also be associated with the detection 
by the cantilever of the path taken by the electron.

The thermally averaged elastic transmission is independent of the 
basis in which the average is performed, but when we examine the terms that
contribute to the average, the balance between dephasing and thermal smearing 
varies considerably.
In the Fock basis, the phase shift induced by the cantilever does not vary
significantly with either the particular state, $|n \rangle$, or the coupling, 
as illustrated in fig 1a. Hence an analysis in the Fock basis
predicts that the thermally averaged  
fringe visibility is largely reduced by dephasing. However, in the coherent 
state basis, the phase shift in the transmission amplitude 
depends strongly on the exact value of the 
coherent state, $|\nu\rangle$. This is illustrated in fig 1b, using just 
three coherent states as an example, but large variations in phase
between different states is a general feature. Thus an analysis in the
 coherent state basis predicts that thermal smearing 
plays an important r\^{o}le in reducing the thermally
 averaged fringe visibility.

A steady-state theory, in which the cantilever is assumed to be in a thermal 
state, cannot distinguish between thermal smearing and 
dephasing and so cannot make predictions that depend on the cantilever 
basis. However, on theoretical grounds, there is good reason to choose the 
coherent state basis as the physically correct one in which to analyse the 
cantilever-dot interaction. This is because for a cantilever interacting with 
a thermal environment the coherent state basis is in some sense `selected' by 
the interaction with the environment\cite{Zurek}. The selection of the
coherent state basis by the environment is predicted as a result of a 
presumed linear coupling between the cantilever and the environment.
An extension of the theory 
to the time-dependent regime offers the prospect of making predictions about 
current fluctuations in the system which should depend on the choice of 
cantilever basis states. Measurements of the fluctuations and comparison with
 a dynamic theory would then allow the selection of coherent states by the 
 environment to be verified.

\section*{Figure Caption}
Argument of the resonant elastic transmission amplitude 
as a function of $\kappa=\lambda/\hbar\omega$ for (a) the Fock states 
$n=1,10$ and $100$, (upper panel), and (b)  the coherent states
$\nu=(1/\sqrt{2})(1-i),\sqrt{5}(1-i)$ and $\sqrt{50}(1-i)$, (lower panel),
with $\Gamma/\hbar\omega=10$. 
 The magnitudes of the coherent states are chosen
so that the mean number of quanta in each correspond to one of
the Fock states. 
\newpage
\epsfig{file=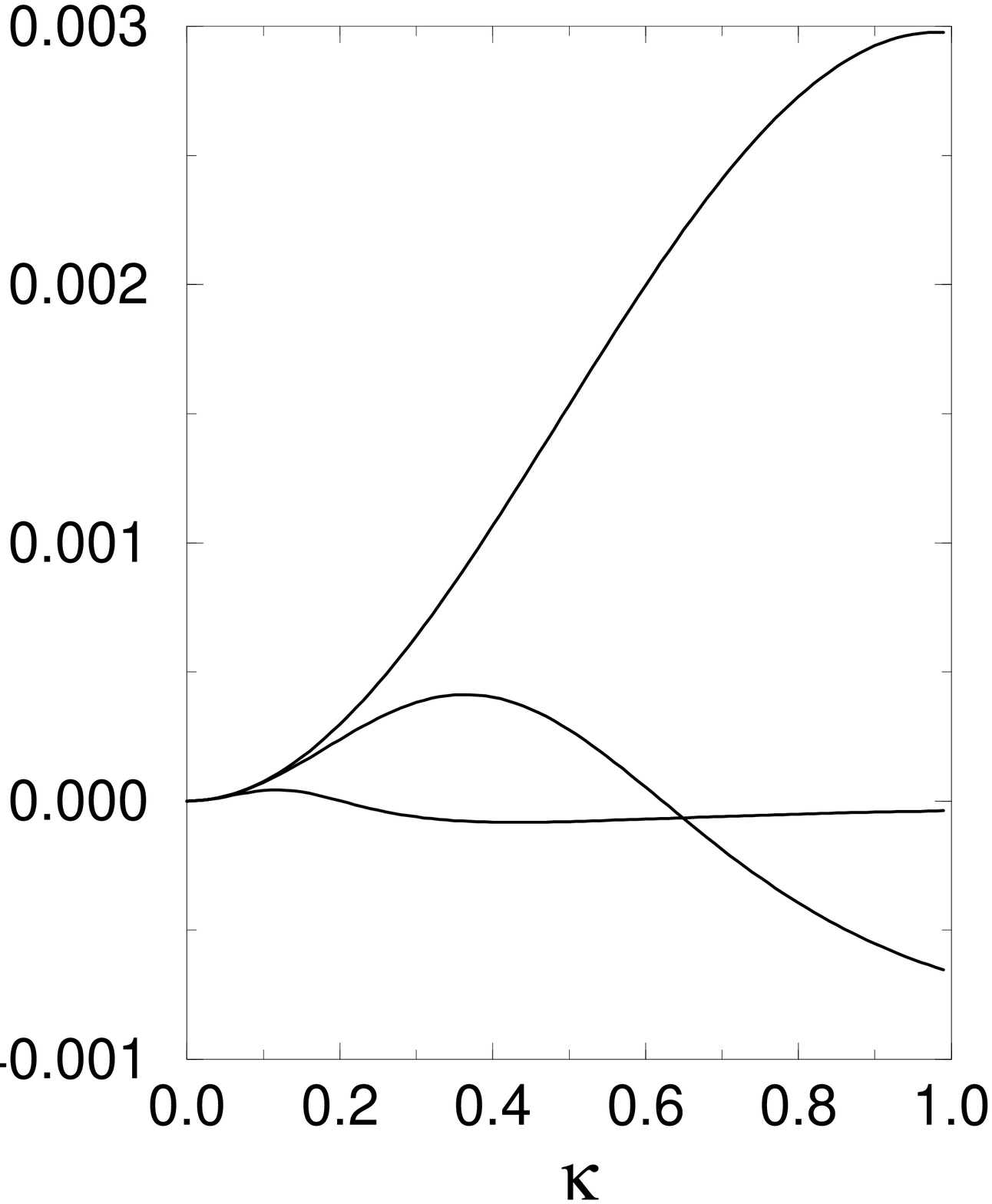, width=5.5cm}

\epsfig{file=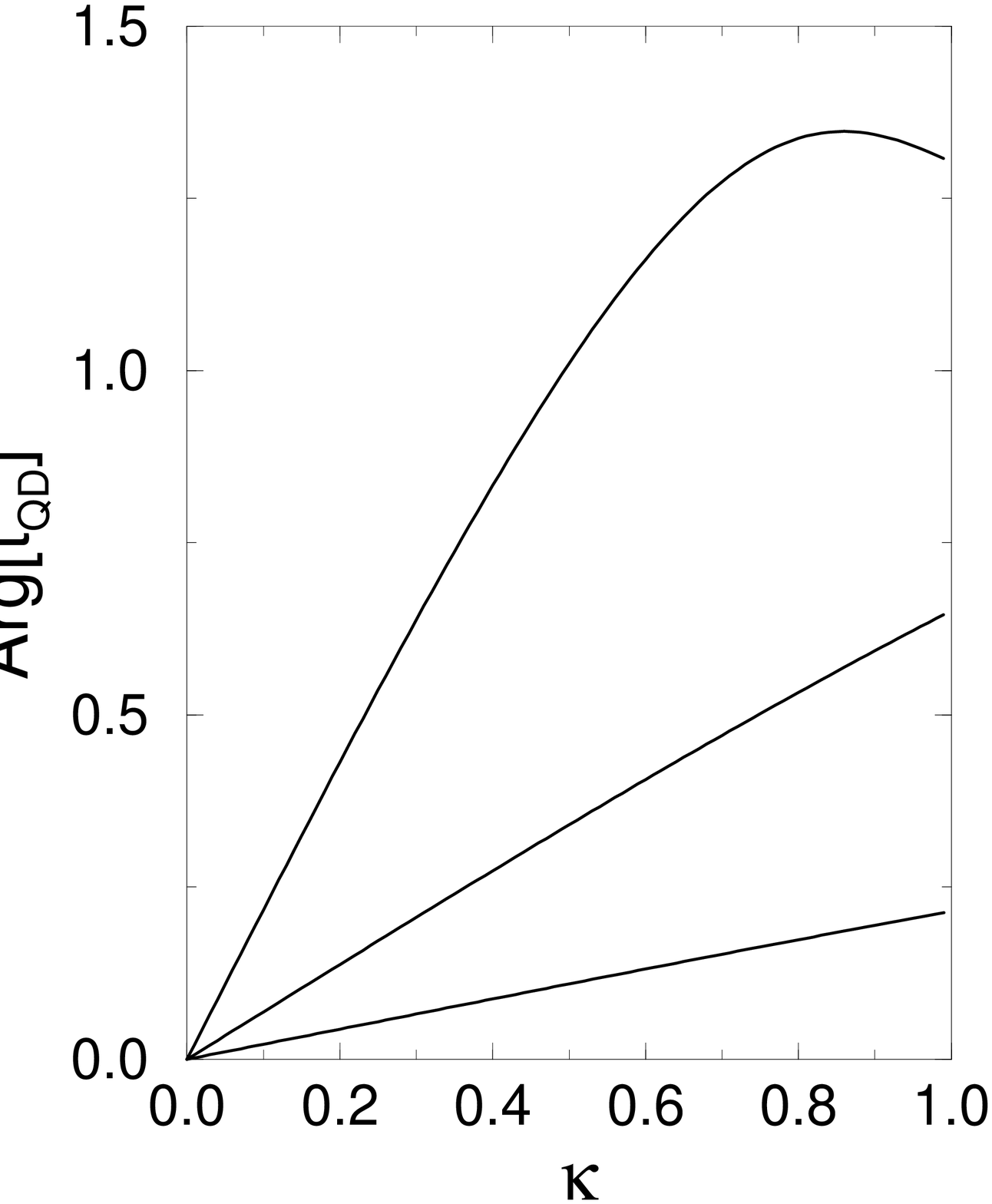, width=5.5cm}
\end{document}